\documentclass{iau}
\usepackage{graphicx,amsmath}

\newcommand{\Msun}{\hbox{$\rm\thinspace \text{M}_{\odot}$}}

\newcommand{\Gyr}{\rm\thinspace \text{Gyr}}

\title[Interpolated stellar oscillation frequencies] 
{Precise and accurate interpolated stellar oscillation frequencies on the main sequence}

\author[Warrick H. Ball et al.]   
{Warrick H. Ball$^{1,2}$, Jesper Schou$^2$, Laurent Gizon$^{2,1}$\\ \and Jo\~ao P. C. Marques$^1$}

\affiliation{$^1$Institut f\"ur Astrophysik, Georg-August-Universit\"at G\"ottingen, \\
Friedrich-Hund-Platz 1, 37077 G\"ottingen, Germany \\ 
email: {\tt wball@astro.physik.uni-goettingen.de} \\[\affilskip]
$^2$Max-Planck-Institut f\"ur Sonnensystemforschung, Max-Planck-Str. 2, \\
37191 Katlenburg-Lindau, Germany}

\pubyear{2013}
\volume{301}
\pagerange{xx--xx}
\jname{Precision Asteroseismology: Celebration of the Scientific Opus of Wojtek Dziembowski}
\editors{W. Chaplin, J. Guzik, G. Handler \& A. Pigulski, eds}

\begin{document}

\maketitle

\begin{abstract}
  High-quality data from space-based observatories present an
  opportunity to fit stellar models to observations of
  individually-identified oscillation frequencies, not just the large
  and small frequency separations.  But such fits require the
  evaluation of a large number of accurate stellar models, which
  remains expensive.  Here, we show that global-mode oscillation
  frequencies interpolated in a grid of stellar models are precise and
  accurate, at least in the neighbourhood of a solar model.

\end{abstract}

\firstsection 

\vskip0.2cm

Asteroseismology from space presents an opportunity to fit stellar
models using sets of individual mode frequencies for a large and
growing number of solar-like oscillators.  This wealth of data can
tightly constrain stellar models but most fitting methods require
either a large number of model evaluations for a single star or a
fixed grid of low resolution to model many stars.  Here, we show that
interpolating in a grid of main-sequence models can provide precise
and accurate stellar oscillation frequencies.  The interpolation is
much faster than the calculation of similarly accurate stellar models
and overcomes deficiency in the grid resolution.  Though stellar model
frequencies have previously been interpolated linearly along
evolutionary tracks (e.g.~\cite[Kallinger et al. 2010]{kallinger10}),
and thus in age, we are unaware of any previous attempts to
interpolate in other stellar model parameters.

We computed a grid of models using the stellar evolution code CESTAM
\cite[(Marques et al. 2013)]{cestam} on a regular grid with 61 ages
$t$ from 3 to 6 Gyr, 11 masses $M$ from 0.975$\Msun$ to 1.025$\Msun$,
11 initial metallicities $Z$ from 0.016 to 0.021, 11 initial hydrogen
contents $X$ from 0.69 to 0.74 and 7 mixing-length parameters $\alpha$
from 1.6 to 1.9.  These were chosen such that the central model is
Sun-like and the range is large enough to characterize parameter
uncertainties.  The oscillation frequencies were calculated with
ADIPLS \cite[(Christensen-Dalsgaard 2008)]{adipls}.  For any set of
parameters within the grid's boundaries, we interpolate the
oscillation frequencies of the corresponding model with cubic splines.

\begin{figure}
  \begin{center}
    \includegraphics[width=0.9\textwidth]{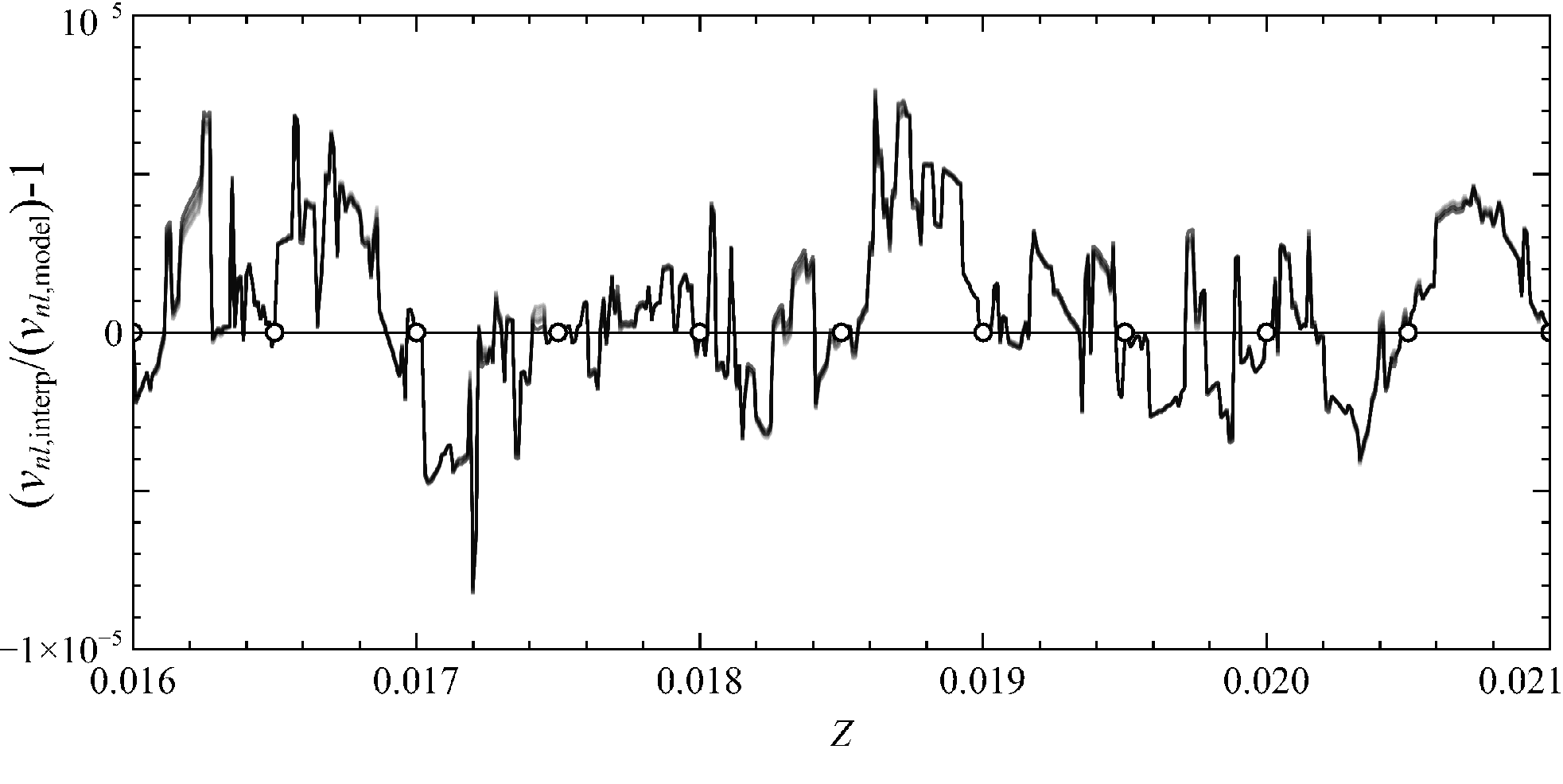}
    \caption{Fractional differences between modelled and interpolated
      frequencies along the metallicity $Z$, with the other parameters
      held fixed at $t=4.5\Gyr$, $M=1\Msun$, $X=0.715$ and
      $\alpha=1.75$.  There are 27 lines plotted, each corresponding
      to a different oscillation mode ($17\leq n\leq25$,
      $0\leq\ell\leq2$) but the lines lie mostly on top of each other
      and cannot be distinguished.  The solid white circles show the
      interpolation points.  The interpolation error is dominated by
      scatter that we attribute to numerical noise.}
    \label{freqi}
  \end{center}
\end{figure}

As a first test of the accuracy of the interpolation, we computed
dense sequences (50 times the grid resolution) of models by taking the
central parameter values and varying one of $M$, $Z$, $X$ or $\alpha$
at a time.  (Age is discussed in the next paragraph.)  The fractional
differences between the oscillation frequencies of these models (for
modes $17\leq n\leq25$, $0\leq\ell\leq2$) for the metallicity $Z$ are
shown in Fig.\,\ref{freqi}.  We note two points.  First, there is
scatter on the order of $10^{-6}$.  We attribute this to the accuracy
of the stellar evolution code, which has a numerical tolerance of
$10^{-6}$ in the mass co-ordinate.  Second, there only appears to be
one curve because the results for 27 different oscillations modes are
plotted over each other.  The error induced by interpolation thus
behaves like a nearly-perfectly correlated fractional error in all the
frequencies.

To model the interpolation error, we divided the grid into two parts.
The first contained models spanning the grid with twice the step in
each parameter.~e.g.~6 metallicities instead of 11.  We then used
these models to interpolate at the parameter values of the other
models and computed the differences in the frequencies between the
stellar models and the interpolated values.  For each frequency, the
fractional errors are approximately normally distributed with the same
scatter of about $4\times10^{-6}$ for all modes.  We assume the errors
are perfectly correlated and construct a covariance for the model
errors.  This is added to the observed covariance matrix when fitting
stellar models to observed frequencies.  Ideally, the interpolation
error would be everywhere much smaller than the observed error.  For
our interpolation routine, the fractional errors correspond to
absolute errors up to about $10$\,nHz, which is several tens times
smaller than frequencies derived from typical \textit{Kepler} or CoRoT
observations.

For the accuracy in the age $t$, we used every fourth model in the
sequence of the central model to interpolate along that model's
evolutionary track.  The output is qualitatively similar to
Fig.\,\ref{freqi} but accurate everywhere to a fractional error
smaller than $10^{-6}$.

Thus, stellar oscillation frequencies can be precisely and accurately
interpolated, provided that the stellar models are themselves
accurately calculated.  This may increase the computational cost of
the grid but not of the interpolation itself.  Finally, the
interpolation should be tested for any given grid and, if
non-negligible, appropriately characterized, noting that the model
errors might be strongly correlated.

The authors acknowledge research funding by Deutsche
Forschungsgemeinschaft (DFG) under grant SFB 963/1 ``Astrophysical
flow instabilities and turbulence'' (Project A18).

\end{document}